\documentclass[sigconf]{acmart}
\usepackage{comment}
\usepackage{xr}
\usepackage[T1]{fontenc}
\usepackage{nameref}
\usepackage{flushend}
\usepackage{multirow,slashbox,pbox,booktabs}
\usepackage[export]{adjustbox}
\usepackage{amsmath,amsfonts,amssymb,amsthm,paralist}
\usepackage{algorithmicx}
\usepackage[normalem]{ulem}
\usepackage[noend]{algpseudocode}
\usepackage[inline]{enumitem}

\makeatletter \let\c@table\c@figure \makeatother

\graphicspath{{../}}

\def\etal{\textit{et al.}}

\copyrightyear{2020}
\acmYear{2020}
\setcopyright{acmcopyright}
\acmConference[CoDS COMAD 2020]{7th ACM IKDD CoDS and 25th COMAD}{January 5--7, 2020}{Hyderabad, India}
\acmBooktitle{7th ACM IKDD CoDS and 25th COMAD (CoDS COMAD 2020), January 5--7, 2020, Hyderabad, India}
\acmPrice{15.00}
\acmDOI{10.1145/3371158.3371199}
\acmISBN{978-1-4503-7738-6/20/01}

\title{Innovation and Revenue: Deep Diving into the\\ Temporal Rank-shifts of Fortune 500 Companies}

\author{Mayank Singh}
\affiliation{%
  \institution{Dept.  of Computer Science and Engg.\\
  IIT Gandhinagar, India}
  }
\email{singh.mayank@iitgn.ac.in}

\author{Arindam Pal}
\affiliation{%
  \institution{Data61, CSIRO
  \\Sydney, NSW, Australia}
}
\email{arindamp@gmail.com}

\author{Lipika Dey}
\affiliation{%
 \institution{TCS Research and Innovation\\
 New Delhi, India}
}
\email{lipika.dey@tcs.com}

\author{Animesh Mukherjee}
\affiliation{%
  \institution{Dept.  of Computer Science and Engg.
  \\IIT Kharagpur, India}
  }
\email{animeshm@cse.iitkgp.ac.in}

\renewcommand{\shortauthors}{Singh et al.}


\hypersetup{draft}
\begin{document}

\begin{abstract}
Research and innovation is an important agenda for any company to remain competitive in the market. The relationship between innovation and revenue is a key metric for companies to decide on the amount to be invested for future research. Two important parameters to evaluate innovation are the quantity and quality of scientific papers and patents. Our work studies the relationship between innovation and patenting activities for several Fortune 500 companies over a period of time. 
We perform a comprehensive study of the patent citation dataset available in the Reed Technology Index collected from the US Patent Office. We observe several interesting relations between parameters like the number of (i) patent applications, (ii) patent grants, (iii) patent citations and Fortune 500 ranks of companies. We also study the trends of these parameters varying over the years and derive causal explanations for these with qualitative and intuitive reasoning. 
To facilitate reproducible research, we make all the  processed patent dataset publicly available.\footnote{https://github.com/mayank4490/Innovation-and-revenue. This research was done, when the second author was working at TCS Research and Innovation}

\end{abstract}

\maketitle
\section{Introduction}

Patent articles contain important research results that are valuable to the industry, academia, business, and policy-making organizations. Patent technology produce novel and industrially usable products which enhance industry's competitive edge. Therefore, industry giants spend extensively in research activities for retaining and increasing their competitive advantages in the corresponding technology groups. Multiple previous works have shown that R\&D outcomes are important assets for any industry giant~\cite{NBERw6984,hall2007market}. World  Intellectual Property Organization (WIPO) reports that nearly 90--95\% of the world's R\&D outcomes are covered in patent publications. Only the remaining 5--10\% are included in the scientific literatures in the form of essays and publications~\cite{liu2008decoding}. Therefore, it is crucial to analyze patent information to understand industrial trends
and compare research growths of several industries in the same technology group.

In this work, we study the effect of research outcomes in the form of patents on revenue generation of top US industry giants. In contrast to previous works, we correlate patenting activities with the Fortune 500 ranks of companies\footnote{Considering ranks instead of revenue values helps to avoid yearwise dollar inflation rates, global economic trends, etc.}, instead of total revenue values. \textbf{The Fortune 500} ($F500$) is an annual ranked list of 500 top-most United States corporations published by the Fortune  magazine
The ranking is based on the total revenue generated in each fiscal year. The lists includes public and private US corporations --- whose revenues are publicly available --- from manufacturing, mining, energy exploration, banking, life insurance, retail, transportation, information technology and service domain. The current list is available at: \url{http://fortune.com/fortune500/list/}.

\noindent\textbf{Limitations of the existing works}: The existing works have several limitations. First, most of these have studied patents granted before 1990s, so they do not consider recent data. Second, existing systems do not consider inter-industry research competition. Third, most of these studies do not take into account the age and field expertise of the companies. Fourth, existing systems utilize crude revenue values that should not be directly compared owing to year-wise dollar inflation rates, global economic trends, etc. 

\noindent\textbf{Our contributions}: We address some of the above limitations in this work by introducing \textit{temporal buckets} that group together companies based on their foundation year and present a thorough correlation study between patenting behavior and performance. 

Towards this objective, we make the following contributions.
\begin{compactitem}
\item We downloaded a massive patent dataset consisting of more than 2.6 million full text patent articles with nearly 93 million patent citations from the Reed Technology Index.
\item We invested extensive manual effort in extraction, cleaning, indexing, and other related preprocessing steps. 
\item We conducted rigorous empirical study on this manually curated dataset by first dividing it into buckets based on the company foundation year. Subsequently, we conduct extensive experiments to identify the correlations between the patenting dynamics of companies and their $F500$ ranks.
\item As a next step, we deep dive further and identify \textit{temporal rank-shifts} of the companies and show that they also correlate well with company R\&D activities.
\item Finally, we further identify that inter-industry citations representing competition could lead to decay/rise in the overall growth of the companies. 
\end{compactitem}

\vspace{-0.2cm}

\section{Related Work}\label{sec:rel_work}
A considerable amount of literature has been published to better understand several parameters like patent citations, number of patent applications, number of patent grants. The first serious discussion and analysis of patent data emerged during the 1990s~\cite{Narin1994}. For better organization, we broadly divide the related work into four subparts: 

\noindent\textbf{General patent analysis}: Derek De Solla Price~\cite{de1969measuring} first showed existence of high positive correlation between scientific output (measured in terms of number of research articles) coming from a country with its gross domestic product (GDP). Two decades later, Narin \etal~\cite{Narin1994} presented evidences of similarity between literature bibliometrics and patent bibliometrics. They showed that the number of granted patents from a country correlates positively with its GDP. James Bessen~\cite{BESSEN2008932} found that patents issued to small patentees are much less valuable than those issued to large corporations. 
Sampat \etal~\cite{Sampat2005} found that citations to research papers are significantly related to the probability that a patent is licensed,  but not to revenues conditional upon licensing. Daim \etal~\cite{DAIM2006981} forecast for three emerging technology areas namely, fuel cell, food safety and optical storage technologies by utilizing of bibliometrics and patent analysis. 

\noindent\textbf{Patents and society}:
Many studies have discussed the potential benefits and negative effects of patenting research on society~\cite{gifford2004social}. Sp{\l}awi{\'n}ski~\cite{splawinski2005patents} has presented an overview of patenting ethics. They present real examples to show how certain patenting activities such as broad claims, poor disclosures, etc., badly affect social benefits and delay technological advancement. 
However, recent trends in patenting activities have led to more openness and existence of fierce competition has resulted in overall technological advancement and an inherent benefit to the society. Tesla, in 2014, made open all of its electric vehicle technology patents to accelerate the advent of sustainable transport and to support open source movement~\cite{tesla}. 

\noindent\textbf{Industry R\&D and market value}: Bronwyn H. Hall conducted several interesting studies to understand market value and patenting output~\cite{NBERw6984,10.2307/1593752,HALL2006971,hall2007market,HALL20101033}. They showed that market value of the manufacturing corporations are strongly related to their knowledge assets and patenting activities beautifully capture this information~\cite{NBERw6984}. In their later work~\cite{10.2307/1593752}, they studied patent and citations between 1963--1995 and claimed that extra citation per patent boosts market value by 3\%. 
They showed that for European corporations also, a firm's Tobin's $q$, defined as the ratio of market value to the replacement value of firm's physical assets, is positively and significantly associated with R\&D and patent stocks~\cite{hall2007market}. In their survey work~\cite{HALL20101033}, they observed that private returns to R\&D are strongly positive and higher than those for ordinary capital. Otto \etal~\cite{OBES:OBES002} studied relationship between innovation and the market value of UK firms in a seven year period (1989--1995). Hsu \etal~\cite{HSU2014116} showed that industries that are more dependent on external finance and that are more high-tech intensive exhibit a disproportionately higher innovation level in countries with better developed equity markets.

\noindent\textbf{Relating $F500$ revenue and patenting activities}: Not much is known about the correlation between Fortune 500 ranks and patenting activities. Wang \etal~\cite{xianwen2010technology} combined social network analysis with the patent co-citation network of Fortune 500 companies to evaluate technology level of an enterprise and also identified their core technical competitive power. In their later work~\cite{Wang2011}, they identified several technology groups based on the co-citation networks. They also studied relationship between leading companies and technology groups. Zhu et al.~\cite{ZHU2000105}  proposed several  diverse measures for characterizing the financial performance of the Fortune 500 companies. Strikingly, they found that only about 3\% companies were operating on the best-practice frontier. 

Unlike most of the previous studies, this paper is different in three aspects -- (i) we present a correlation study between temporal ranks (instead of crude revenue values) of $F500$ companies with patenting outputs,  (ii)  we meticulously overcome the bias of experience/age by introducing three temporal buckets, and (iii) as opposed to previous works, we present empirical evidences that inter-company in-citations correlate well with rise/fall in ranks. 

\section{Dataset}
\label{sec:dataset}
We compile two datasets for the current study. \\
\noindent\textbf{Patent dataset}: 
We construct a structured patent dataset by crawling full text articles indexed in  Reed Technology Index~\cite{reedtech}. The compiled patent dataset consists of patent metadata, such as the unique patent identifier (assigned after the patent granting process), the application year, the grant year, the patent title, the applicants' name, the company name, etc.,  along with patent bibliography, such as the patent citations, the non-patent citations including scientific citations, urls, blogs, white papers, etc. Table~\ref{tab:dataset} presents the detailed description of the compiled dataset. The processed dataset is available at: \url{https://github.com/mayank4490/Innovation-and-revenue}.
 
\begin{table}[t]

\centering \normalsize
  \caption{General statistics of the compiled patent dataset.}
  \label{tab:dataset}
  \vspace{-0.2cm}
  \small{
  \begin{tabular}{lc} \toprule
  Patent count & 2,608,782\\
  Grant year range & 2005--2017 \\
  Application year range  & 1965--2016  \\ 
  Number of citations & 93,938,858 \\
  Number of patent citations & 75,118,567 \\
  Number of non-patent citations &  18,820,291 \\
  \bottomrule \hline
 \end{tabular}}
 \vspace{-0.5cm}
\end{table}

\begin{figure}[!tbh]
\vspace{-0.5cm}
 \centering
 \includegraphics[width=0.7\hsize]{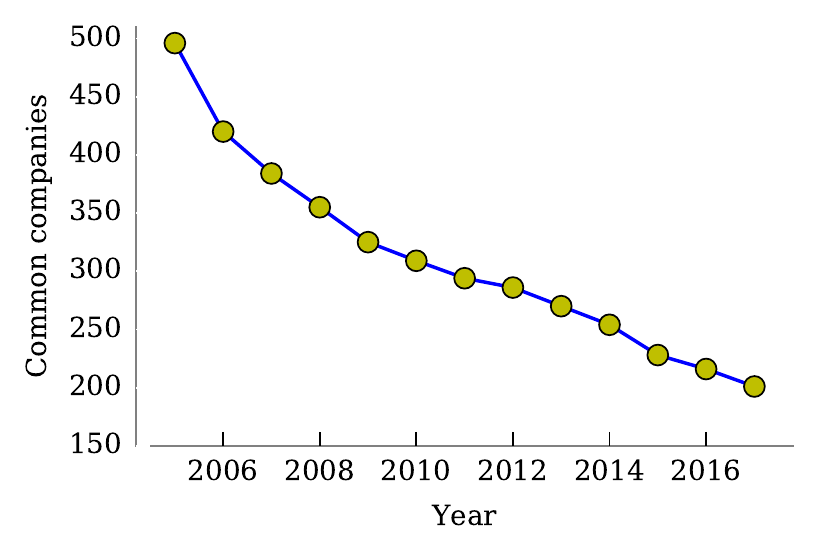}
   \vspace{-0.5cm}
  \caption{Decay in the number of common companies with the inclusion of the newer rank lists over the years. 
  }
\label{fig:decay_count}
\vspace{-0.6cm}
\end{figure}

\noindent\textbf{F500 dataset}: We compile the Fortune 500 rank lists published between 2005--2017. The major challenge in the compilation process was to normalize the company names present in different lists. Figure~\ref{fig:decay_count} shows the decay in the number of common companies as newer yearly rank lists are taken into consideration. Overall, we find 201 companies that are present across all the rank lists.

\section{50-year temporal buckets}
\label{sec:temp_buck}
In this section, we detail the construction procedure of the 50-year \textit{temporal buckets}. 

\noindent\textbf{Filtering}: Out of the 201 companies in $F500$ dataset, only 72 companies have at least 100 patents granted between 2005--2017. We also discard few very old (before 1850) and
new (after 2000) companies to remove ``corner'' cases. Overall, we find 68 companies that satisfies all the above criterion. The rest of the paper presents all the experiments on these 68 companies. 

\noindent\textbf{Bucketing}: We divide these 68 companies into three 50-year \textit{temporal buckets} to perform the subsequent experiments. The first bucket (\textbf{bucket I}) consists of companies founded between 1851--1900. The \textbf{bucket II} and \textbf{bucket III} consists of companies founded between 1901--1950 and 1951--2000 respectively. The proposed bucketing scheme eliminates the normalization efforts needed to accommodate the company age. Buckets I, II and III consists of
16, 29 and 23 companies respectively (see Table~\ref{tab:company_list}). Bucket I mostly consists of consumer product companies, while bucket III mostly comprises information technology companies. Bucket II consists of a mixture of several groups including consumer product, information technology and automobiles. 

\begin{table*}[!tbh]
\centering
\caption{Buckets I, II and III consist of companies founded between 1851--1900, 1901--1950 and 1951--2000 respectively.
}
\vspace{-0.2cm}
\label{tab:company_list}
\small{
\begin{tabular}{lcclcclc} \toprule
\multicolumn{2}{c}{\textit{Bucket I}} &&\multicolumn{2}{c}{\textit{Bucket II}}&&\multicolumn{2}{c}{\textit{Bucket III}}\\\cline{1-2} \cline{4-5} \cline{7-8}
\multicolumn{1}{l}{ \textit{\small Company name}}  & \textit{\small Foundation year} &&\multicolumn{1}{l}{ \textit{\small Company name}}  & \textit{\small Foundation year}&&\multicolumn{1}{l}{ \textit{\small Company name}}  & \textit{\small Foundation year}\\
\midrule
Corning              & 1851 && Archer Daniels Midland          & 1902 && Comcast                & 1963 \\
General Mills        & 1856 && Ford Motor                      & 1903 && Nike                   & 1964 \\
Kimberly Clark       & 1872 && Harley Davidson                 & 1903 && Applied Materials      & 1967 \\
Conocophillips       & 1875 && Rockwell Automation             & 1903 && Quest Diagnostics      & 1967 \\
Ball                 & 1880 && Honeywell International         & 1906 && Intel                  & 1968 \\
PPG Industries       & 1883 && Kellogg                         & 1906 && First Data             & 1971 \\
Avon Products        & 1886 && Xerox                           & 1906 && Microsoft              & 1975 \\
Johnson \& Johnson      & 1886 && Baker Hughes                 & 1907 && Oracle                 & 1977 \\
Bristol Myers Squibb & 1887 && General Motors                  & 1908 && Micron Technology      & 1978 \\
Abbott Laboratories  & 1888 && IBM 			       & 1911 && Boston Scientific      & 1979 \\
Emerson Electric     & 1890 && Whirlpool                       & 1911 && AT\&T                   & 1983 \\
General Electric     & 1892 && Illinois Tool Works             & 1912 && Cisco Systems          & 1984 \\
International Paper  & 1898 && Lear                            & 1917 && Qualcomm               & 1985 \\
Pepsico              & 1898 && Parker Hannifin                 & 1917 && Staples                & 1986 \\
General Dynamics     & 1899 && Cummins                         & 1919 && Capital One Financial  & 1988 \\
Weyerhaeuser         & 1900 && Halliburton                     & 1919 && Agco                   & 1990 \\
\multicolumn{2}{c}{} 		&& Eastman Chemical                       & 1920 && Time Warner            & 1990 \\
\multicolumn{2}{c}{}  && Raytheon                              & 1922 && Amazon             & 1994 \\
\multicolumn{2}{c}{}  && Ecolab                                & 1923 && Northrop Grumman       & 1994 \\
\multicolumn{2}{c}{} && Textron                         	& 1923 && Lockheed Martin        & 1995 \\
\multicolumn{2}{c}{} && Caterpillar                  	  	 & 1925 && Autoliv                & 1997 \\
\multicolumn{2}{c}{}  && Masco                         		 & 1929 && Monsanto               & 2000 \\
\multicolumn{2}{c}{}  && Texas Instruments          	  	   & 1930 && Verizon Communications & 2000 \\
\multicolumn{2}{c}{}  && Baxter International       	  	   & 1931 && \multicolumn{2}{c}{} \\
\multicolumn{2}{c}{}  && Morgan Stanley             		     & 1935 && \multicolumn{2}{c}{} \\
\multicolumn{2}{c}{}  && Hewlett Packard          	       & 1939 &&  \multicolumn{2}{c}{}  \\
\multicolumn{2}{c}{}  && Nucor                     	      & 1940 &&  \multicolumn{2}{c}{}   \\
\multicolumn{2}{c}{}   && Mattel                  	        & 1945 &&   \multicolumn{2}{c}{}    \\
\multicolumn{2}{c}{}  && Stryker                  	       & 1946 &&\multicolumn{2}{c}{}    \\\bottomrule
\hline
\end{tabular}}
\vspace{-0.2cm}
\end{table*}

\noindent\textbf{Name normalization}: We find huge variations in the company names within the patent metadata. These variations exist due to industry organization hierarchy such as different geographic locations of the research labs, several technology teams, collaborations, etc. We also find huge number of variations resulting from spelling errors, acronyms, etc. Table~\ref{tab:normalized_instances} shows one representative example. We manually normalize the different names of all the 68 companies that we experiment on. AT\&T has the maximum number of unnormalized variations (total 95).

\begin{table*}[!tbh]
\centering
\caption{Representative company showing respective unnormalized instances occurring in the patent dataset. 
}\label{tab:normalized_instances}
\vspace{-0.2cm}
\small{
\begin{tabular}{ll} \toprule
Normalized name & Unnormalized variations \\
\midrule

Stryker & \parbox[t]{14cm}{stryker development llc, stryker biotech, stryker canadian management, stryker combo l l c, stryker coropration, stryker endoscopy, stryker ireland, stryker endo, stryker european holdings i llc, stryker france, stryker gi services c v, stryker  stryker gi, stryker, stryker instruments stryker leibinger gmbh co kg, stryker leibinger gmbh co kg, stryker nv operations, stryker ortho pedics, stryker orthopaedics, safe orthopaedics, stryker puerto rico, stryker trauma ag, stryker trauma gmbh, stryker truama s a, stryker trauma sa, stryker trauma s a, stryker spine, styker spine}\\

\bottomrule
\hline
\end{tabular}}
\vspace{-0.2cm}
\end{table*}

\section{Correlating temporal ranks and patenting patterns}
\label{sec:corr_exp}

\subsection{Experimental setup}
\label{sec:exp_setup}
We employ standard Pearson's correlation coefficient metric~\cite{lee1988thirteen} for computing the correlation between companies' patenting activity parameters (e.g., grant count, application count, citation count, etc.) and the respective $F500$ ranks. The next four experiments are categorized into two sets as follows:
\begin{enumerate}[noitemsep,nosep]
\item Correlating current patenting activities with next five year $F500$ ranks. The two experiments are described in Section~\ref{sec:cor_exp1} and Section~\ref{sec:cor_exp3}.
\item Correlating current $F500$ ranks with the next five year patenting activities. The two experiments are described in Section~\ref{sec:cor_exp2} and Section~\ref{sec:cor_exp4}.  
\end{enumerate}

\vspace{-0.4cm}
\subsection{Effect of patenting on \textit{F500} ranks}
\begin{figure}[b]
\vspace{-0.5cm}
 \centering\includegraphics[width=0.9\hsize]{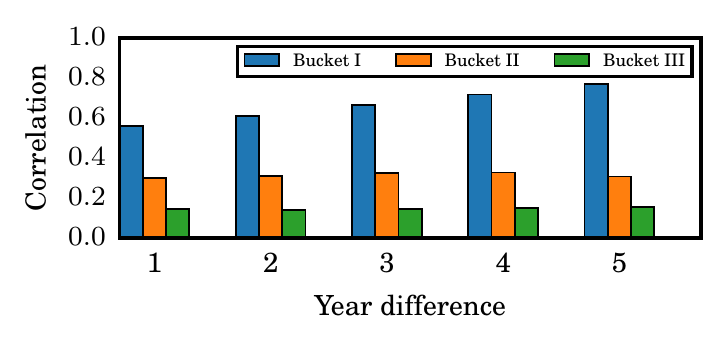}
   \vspace{-0.5cm}
  \caption{(Best viewed in color) Average correlation between patent grant count drawn from seven different start years and $F500$ ranks of the next five years, $\delta = \{1,2,3,4,5\}$. 
  }
  \label{fig:average_bucket_patent_grant}
  \vspace{-0.5cm}
\end{figure}

In this experiment, we measure the correlation between the `current' patent grant count of a company with its next five year (denoted by $\delta$) $F500$ ranks. For the 68 companies in our list, we draw the `current' grant count from seven different years (2005--2011), call these as `start' years and estimate the correlation of each of these start year with the $F500$ ranks of the next five years.   Figure~\ref{fig:average_bucket_patent_grant} 
illustrates the average correlation over the seven different start years. Each bucket shows a different temporal 
characteristics. 

\noindent\textbf{Key observations}: We observe higher correlation values for bucket I compared to the buckets II and III. Bucket I companies' future revenue is therefore heavily dependent on the current patenting volume. For all the three buckets, an overall positive correlation indicates that companies with higher patenting volume tend to garner higher revenues. A further interesting point is that for bucket I companies, the effect of current patenting volume is more pronounced on the revenue garnered in the later years demonstrated by the overall increase in the correlation value. However, the correlation seems to remain stable for the two other buckets. 

\label{sec:cor_exp1}

\begin{figure}[t]
\vspace{-0.3cm}
 \centering\includegraphics[width=0.9\hsize]{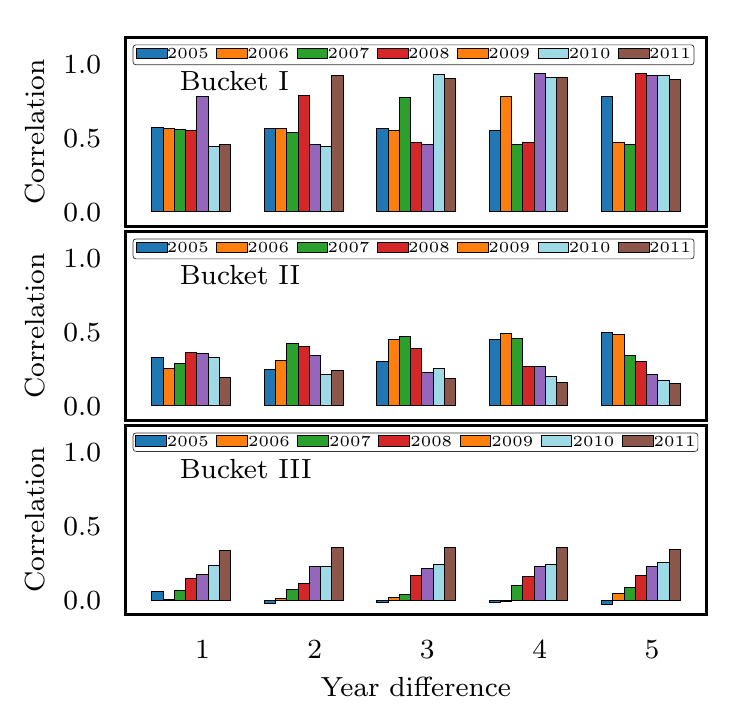}
 \vspace{-0.5cm}
  \caption{(Best viewed in color) Correlation between patent grant count and future $F500$ rank at five consecutive years, $\delta = \{1,2,3,4,5\}$, for seven different start years, 2005--2017. 
  }
  \label{fig:grant_rank_cor}
  \vspace{-0.4cm}
\end{figure}

\noindent\textbf{More experiments}: We take a step further, reporting in Figure~\ref{fig:grant_rank_cor}, the correlation values for each of the seven start years separately. Similar to our earlier observations, for each start year, we find a higher correlation for bucket I as compared to the buckets II and III. Bucket I shows that the correlation is above 0.8 for majority (4 out of 7) of the start years at $\delta = 5$. This leads us to claim that for the bucket I companies, the patenting volume affects the $F500$ ranks more sharply in the long run.

In Figure~\ref{fig:grant_rank_cor}, bucket II shows an interesting trend. Initial start years exhibit a higher correlation as compared to subsequent start years. Interestingly, for the last two start years (2010 and 2011) the correlation is quite low for all the different $\delta$. This leads us to conclude that the dependence of company revenue on the patenting volume is on a steady decline for the bucket II companies.  

Lastly, bucket III shows same trends for every individual value of $\delta$. The two key observations here are: (i) correlation remains invariant for different $\delta$ values, and (ii) as opposed to bucket II, initial start years exhibit a significantly low correlation as compared to subsequent start years. Therefore, for this bucket, as time progresses, there is a steady rise in the influence of patenting volume on the future $F500$ ranks. 

\subsection{Effect of \textit{F500} ranks on patenting}
\label{sec:cor_exp2}
\begin{figure}[!tbh]
\vspace{-0.05cm}
 \centering\includegraphics[width=0.9\hsize]{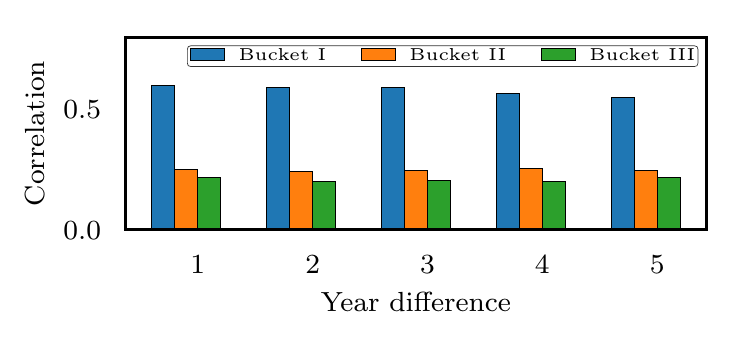}
 \vspace{-0.5cm}
  \caption{(Best viewed in color) Correlation between $F500$ rank and future patent application count averaged over seven different start years for five consecutive years, $\delta = \{1,2,3,4,5\}$. 
  }
  \label{fig:average_bucket_appl_rank}
  \vspace{-0.5cm}
\end{figure}

In this section, we perform the reverse experiment. We correlate the `current' $F500$ rank of the companies with their respective patent application counts in the next five years (denoted by $\delta$). Once again, for the 68 companies in our list, we draw the `current' $F500$ ranks from seven different years (2005--2011), call these as `start' years and estimate the correlation of each of these start year with the respective patent application counts in the next five years. Figure~\ref{fig:average_bucket_appl_rank} illustrates this correlation by averaging over seven different start years (2005--2011). Similar to Figure~\ref{fig:average_bucket_patent_grant}, here also, each bucket shows a different temporal characteristic.

\begin{figure}[!tbh]
\vspace{-0.05cm}
 \centering\includegraphics[width=0.9\hsize]{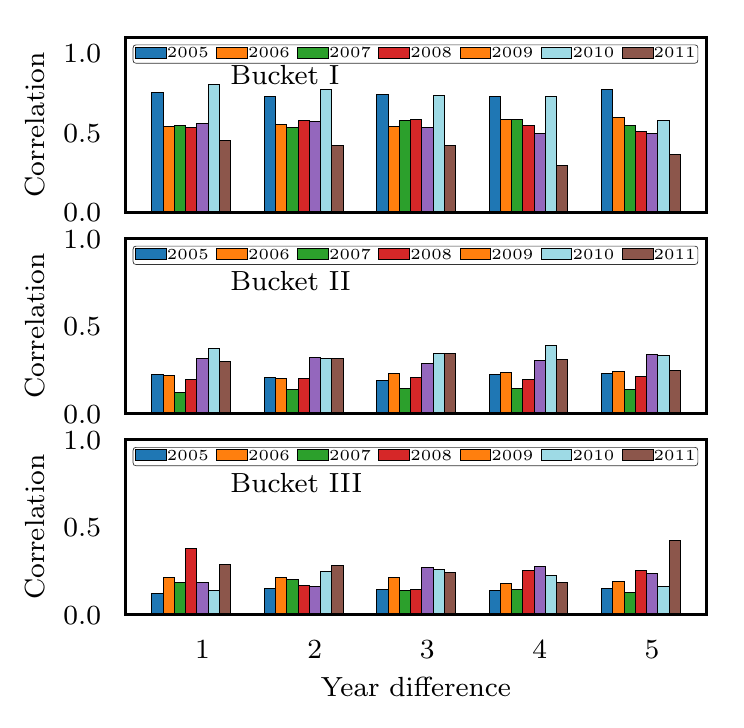}
 \vspace{-0.5cm}
  \caption{(Best viewed in color) Correlation between $F500$ rank and future application count in the five consecutive years, $\delta = \{1,2,3,4,5\}$, for seven different start years, 2005--2011.}
  \label{fig:rank_application_cor}
  \vspace{-0.5cm}
  \end{figure}

\noindent\textbf{Key observations}: We observe higher correlation values for bucket I compared to buckets II and III. For bucket I companies, current revenue seems to strongly drive future patenting volume. Companies with better ranks tend to produce higher overall research output and vice-versa. In Figure~\ref{fig:rank_application_cor}, we present correlation for each start year separately. All the three buckets exhibit a low correlation during the ``global recession'' period (2007--2009)~\cite{wiki}. 

\vspace{-0.4cm}
\subsection{Effect of incoming citations on \textit{F500} ranks}
\label{sec:cor_exp3}
This experiment is similar to the one outlined in Section~\ref{sec:cor_exp1} except that the grant count is replaced by the overall incoming citations to all the patents produced by the company. Interestingly, we observe very similar trends as noted in Section~\ref{sec:cor_exp1} (figure not shown). We observe higher correlation values for bucket I compared to the buckets II and III. Bucket I companies' future revenue is therefore heavily dependent on the current incoming citation volume.

\vspace{-0.3cm}
\subsection{Effect of \textit{F500} ranks on incoming citations}
\label{sec:cor_exp4}
This experiment is a similar to the one discussed in Section~\ref{sec:cor_exp2} except that here the next five year application count is replaced by the next five year incoming citations to all the patents produced by the company. The results exhibit very similar trends as those in Section~\ref{sec:cor_exp2} (figure not shown). We observe higher correlation values for bucket I compared to bucket II and III. For bucket I companies, current revenue strongly drive future incoming citation volume.

\vspace{-0.2cm}
\subsection{Possible explanations}
\begin{figure}[!tbh]
 \centering\includegraphics[width=0.8\hsize]{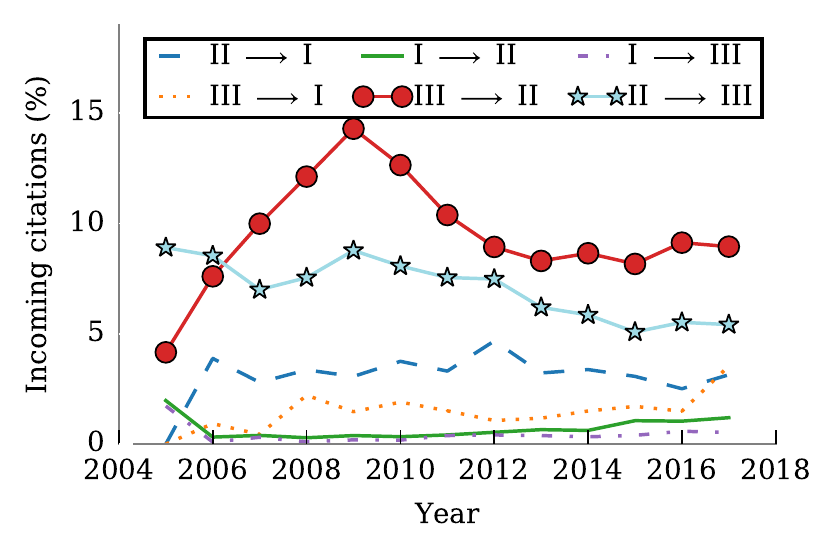}
 \vspace{-0.4cm}
  \caption{Proportion of incoming citations. 
  }
  \label{fig:interbucket_citations}
  \vspace{-0.4cm}
\end{figure}

In this section, we attempt to explain the overall observations that we made in the last four sections. 
In particular, for each bucket, we study the incoming citations from the other two buckets between 2005--2017. Figure~\ref{fig:interbucket_citations} shows the yearwise proportion of inter-bucket incoming citations. As can be noted, bucket I receives marginal number of incoming citations from both bucket II and III. Bucket II receives less incoming citations from bucket I but a considerably large number of incoming citations from bucket III. Bucket III, on the other hand, receives less incoming citations from bucket I and a moderate volume of incoming citations from bucket II. A crucial point to stress here is that the volume of incoming citations from bucket III to bucket II is much larger compared to the other direction (i.e., bucket II to bucket III). We term this as a form of \textit{knowledge stealing}, i.e, bucket III is able to `steal' many more novel ideas from bucket II and build up on them than the other way round. Bucket I is self sustained, witnesses least competition, neither cites nor receives high volume of incoming citations from the rest of the two buckets. Note that Figure~\ref{fig:interbucket_citations} do not show  proportion of self-citations and citations coming from the rest of the companies not considered in this study. A natural analogy is that the bucket I companies behave like `cocoons', the bucket II companies behave like `larva' and the bucket III companies behave like `butterflies'.

We, next, select two representative companies from each of the three buckets and list the top 10 highly citing companies (see Table~\ref{tab:top_incoming_citations}). Once again, it is apparent that a considerable fraction of incoming citations to bucket II companies arrive from bucket III. In contrast, bucket I companies have most of the incoming citations coming from the same bucket itself. 

\begin{table}[t]
 
\footnotesize
\centering
\caption{Inter-company incoming citations: Two representative companies from each bucket showing top 10 citing companies. Values in parenthesis indicate percentage of incoming citations from each citing company.}
\label{tab:top_incoming_citations}
\begin{tabular}{ll}\toprule
\multicolumn{2}{c}{\textbf{\hspace{-1cm} Bucket I: the cocoon}} \\
\textit{Johnson \& Johnson}&\textit{Pepsico}\\\cline{1-2}
Johnson \& Johnson (14.3) & The Coca-Cola (27.5)	\\
Abbott Laboratories (11.0) & Pepsico (17.9)		\\
Novartis (10.7)		  & Meadwestvaco (8.3)		\\
Brien Holden Vision Inst. (6.9)& Concentrate Mfg. (2.5)	\\
Pixeloptics (2.6)	& Kimberly Clark (1.7)		\\
Coopervision Int. (2.3)	& Food Equipment Tech. (1.7)		\\
Google (2.3)		& Crestovo (1.7)		\\
Mcneil (2.1)		& Givaudan (1.7)		\\
The Procter \& Gamble (1.6)& Bunn-o-matic (1.7)		\\
E-vision (1.5)		& Starbucks (1.7)		\\\toprule
\multicolumn{2}{c}{\textbf{\hspace{-1cm} Bucket II: the larva}} \\
\textit{IBM}&\textit{Hewlett Packard}\\\cline{1-2}
IBM (23.5) 		& IBM (8.5)			\\
Microsoft (6.2)	& Hewlett Packard (6.5)	\\
Google (2.2)		& Semiconductor Energy Lab. (5.7)	\\
Apple (1.9)	& Microsoft (4.5)  	\\
Oracle (1.7)		& Google (2.0)		\\
Taiwan Semiconductor (1.4)& Qualcomm (1.9)		\\
Intel (1.4)		& Apple (1.7)			\\
Tela Innovation (1.4)	& Intel (1.6)			\\
Micron Technology (1.3)& Canon (1.4)		\\
Hewlett Packard (1.1)	& Samsung (1.1)		\\\toprule
\multicolumn{2}{c}{\textbf{\hspace{-1cm} Bucket III: the butterfly}} \\
\textit{Intel}&\textit{Microsoft}\\\cline{1-2}
Intel	(14.3)				& Microsoft (20.1)\\	
IBM	(9.0)				& IBM (6.1)\\	
Taiwan Semiconductor (3.7)	& Apple (4.7)\\	
Microsoft	(3.1)			& Google (4.0)\\	
Micron Technology	(2.5)			& Oracle (1.4)\\	
Qualcomm	(2.2)		& Amazon (1.4)\\	
Samsung	(1.7)			& AT\&T (1.2)\\	
Apple 	(1.5)			& Qualcomm (1.0)\\	
United Microelectronics	(1.5)	& Samsung (0.9)\\	
Google	(1.2)				& SAP (0.9)\\	

\bottomrule \hline     
\end{tabular}
\vspace{-0.7cm}
\end{table}

\begin{figure*}[!tbh]
\vspace{-0.3cm}
\centering\includegraphics[width=0.95\hsize]{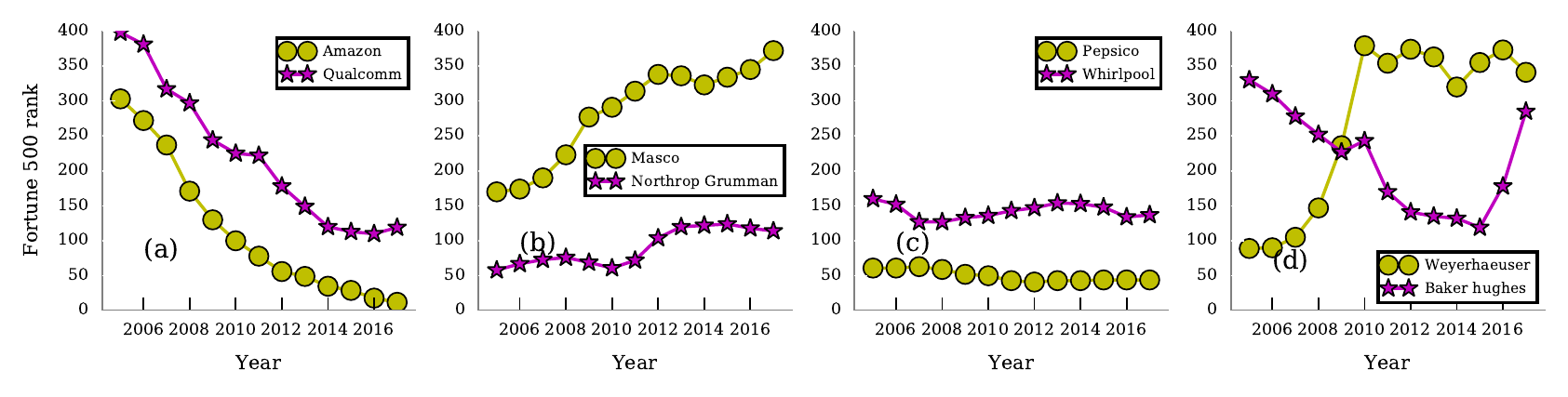}
\vspace{-0.7cm}
  \caption{Two representative companies from each of the four categories. a) \textit{MonInc}, b) \textit{MonDec}, c) \textit{Stable}, d) \textit{Others}. Note that when the ranks are decreasing, the companies are actually rising towards the top (and vice-versa).}
  \label{fig:four_category}
  \vspace{-0.3cm}
\end{figure*}

\section{Temporal rank-shifts}
\label{sec:temp_rank_shift}
In this section, we consider a 13-year time period to understand the shifts in the $F500$ rank profiles. 68 companies are classified into four broad categories based on their rank-shift profiles\footnote{We perform 5-year moving average to make the rank-shift profiles smooth.}:
\begin{enumerate}
 \item \textbf{MonInc}: Rank of a company is becoming better monotonically over (lower and lower) time. The revenue for these companies is therefore on a rise as time progresses (hence the name \textit{MonInc}). At least 80\% of the consecutive year ranks differences are positive. 
 \item \textbf{MonDec}: Rank of a company is worsening monotonically (higher and higher) over time. The revenue for these companies is therefore going down as time progresses (hence the name \textit{MonDec}). At least 80\% of the consecutive year ranks differences are negative. 
 \item \textbf{Stable}: Rank of a company remains stable over the years. This classification is carried out in two steps; first, we compute 13-year average ($mean$) of ranks. Next, we select companies having at least 80\% year ranks between $mean \pm 10$. We also experiment with other variations like $mean \pm stdev$, $mean \pm 2*stdev$, etc. The one we have chosen is able to produce the most clear separation from the other categories.
\item \textbf{Others}: Remaining companies are kept in this category. This category includes companies with rank profiles having multiple crests and troughs over time.
\end{enumerate}

\noindent\textbf{Key observations}: Overall, we find 14, 5, 23 and 26 companies in \textit{MonInc}, \textit{MonDec}, \textit{Stable} and \textit{Others} categories respectively. Figure~\ref{fig:four_category} shows the rank profiles of two representative companies from each category. 
Table~\ref{tab:cat_prof} groups together companies from different buckets in each category. It utilizes a color scheme to represent each bucket -- \textcolor{green}{green} color for bucket I, \textcolor{red}{red} color for bucket II and \textcolor{blue}{blue} color for bucket III. Interestingly, majority of the items in \textit{MonInc} are from bucket III, i.e., the `butterfly' companies. This observation once again reinforces our earlier claim that these companies are improving upon their ranks by drawing knowledge (i.e., `stealing') from bucket II companies and effectively building newer and more innovative ideas on them. In contrast, \textit{MonDec} consists of majority of bucket II, i.e. the `larva' companies. This indicates that the bucket II companies are not able to draw and build up on the knowledge generated in the other buckets to improve upon their ranks. This might be a potential sign of such companies `drying up' in the near future.

\begin{table}[!tbh]
\footnotesize
\centering
\caption{(Best viewed in color) List of companies in each rank-shift profile category. \textcolor{green}{Green}, \textcolor{red}{red} and \textcolor{blue}{blue} color represent buckets I, II and III respectively. }
\label{tab:cat_prof}
\begin{tabular}{lll}\toprule
\textbf{MonInc}          		& \textbf{MonDec}      				& \textbf{Stable}   \\\midrule
\textcolor{green}{Corning} 		& \textcolor{green}{Bristol Myers Squibb}    	& \textcolor{green}{Kimberly Clark}  \\
\textcolor{green}{General Mills}        & \textcolor{red}{Texas Instruments}     	& \textcolor{green}{Johnson Johnson} \\
\textcolor{red}{Ecolab}           	& \textcolor{red}{Masco}       			& \textcolor{green}{Emerson electric}\\
\textcolor{red}{Stryker}                & \textcolor{red}{Morgan Stanley}               & \textcolor{green}{General Electric}\\
\textcolor{blue}{Amazon}		& \textcolor{blue}{Northrop Grumman} 		& \textcolor{green}{Pepsico}   \\
\textcolor{blue}{Comcast} 		&                      				& \textcolor{green}{General dynamics}\\
\textcolor{blue}{Capital One Financial} &                      				& \textcolor{red}{Archer Daniels Midland}          \\
\textcolor{blue}{Micron Technology}     &                     				& \textcolor{red}{Ford Motor}    \\
\textcolor{blue}{Autoliv}		&                      				& \textcolor{red}{Honeywell International}         \\
\textcolor{blue}{Qualcomm}              &                      				& \textcolor{red}{Xerox}         \\
\textcolor{blue}{Monsanto}              &                      				& \textcolor{red}{General Motors}   \\
\textcolor{blue}{Cummins}               &                      				& \textcolor{red}{IBM}            \\
\textcolor{blue}{Oracle}             	&                      				& \textcolor{red}{Hewlett Packard}   \\
\textcolor{blue}{Nike}              	&                      				& \textcolor{red}{Whirlpool}      \\
					&                      				& \textcolor{red}{Nucor}           \\
					&                     				&\textcolor{red}{ Caterpillar}     \\
					&                      				& \textcolor{blue}{Intel}          \\
					&                      				& \textcolor{blue}{AT\&T}          \\
					&                      				& \textcolor{blue}{Microsoft}     \\
					&                      				& \textcolor{blue}{Cisco Systems} \\
					&                      				& \textcolor{blue}{Staples}         \\
					&                      				& \textcolor{blue}{Verizon Communications}          \\
					&                      				& \textcolor{blue}{Lockheed Martin}  \\
\bottomrule
\hline         
\end{tabular}
\vspace{-0.5cm}
\end{table}

\section{Conclusion}
\label{sec:end}
We conduct the first plausible correlation study between research output with the Fortune 500 ranks. 
An interesting future direction would be to automatically predict future revenue of companies based on the correlations established here.
\clearpage
\bibliographystyle{ACM-Reference-Format}
\bibliography{sigproc}
\flushend
\end{document}